\documentstyle[12pt,aaspp4]{article}
\renewcommand{\vec}[1]{\mbox{$\bf #1$}}
\newcommand{\beq}{\begin{equation}}
\newcommand{\eeq}{\end{equation}}

\newcommand{\kms}{\mbox{km s$^{-1}$}}

\newcommand{\cc}{\mbox{ cm$^{-3}$}}

\newcommand{\vecv}{{\bf v}}
\newcommand{\vecb}{{\bf B}}
\newcommand{\tf}{\mbox{$t_{flow}$}}
\newcommand{\An}{\mbox{$A_n$}}

\begin{document}
\title{Nonlinear Development and Observational Consequences of Wardle
C-Shock Instabilities}

\author{Mordecai-Mark Mac Low}
\affil{Max-Planck-Institut f\"ur Astronomie, K\"onigstuhl 17, D-69117
Heidelberg, Germany; mordecai@mpia-hd.mpg.de}

\and

\author{Michael D. Smith} 
\affil{Astronomisches Institut der Universit\"at W\"urzburg, Am Hubland,
D-97074 W\"urzburg, Germany; smith@astro.uni-wuerzburg.de}

\author{ApJ, submitted, 26 March 1997}
\slugcomment{ApJ, submitted, 26 March 1997}

\begin{abstract}

We compute the nonlinear development of the instabilities in C-shocks
first described by Wardle, using a version of the ZEUS code modified
to include a semi-implicit treatment of ambipolar diffusion.  We find
that, in three dimensions, thin sheets parallel to the shock velocity
and perpendicular to the magnetic field lines form.  High resolution,
two-dimensional models show that the sheets are confined by the
Brandenburg \& Zweibel ambipolar diffusion singularity, forcing them
to numerically unresolvable thinness.  Hot and cold regions form
around these filaments, disrupting the uniform temperature structure
characteristic of a steady-state C-shock. This filamentary region
steadily grows as the shock progresses.  We compare steady-state to
unstable C-shocks, showing excitation diagrams, line ratios, and line
profiles for molecular hydrogen lines visible in the K-band, with the
{\em Infrared Space Observatory}, and with NICMOS on the {\em Hubble
Space Telescope}.

\end{abstract}

\keywords{}

\clearpage
\section{Introduction}

Within molecular clouds, interstellar gas becomes dense enough to
recombine almost completely, reaching levels of fractional ionization
less than $\chi = 10^{-4}$.  Positively charged ions or grains carry
the magnetic field in most regions except the central portions of
protostellar disks, where neutral number densities exceed $10^{10}$
cm$^{-3}$ and electron-ion drift becomes important ({\em e.g.}
K\"onigl 1989) \markcite{kon89}.  The charged particles carrying the
field couple in turn with the neutral gas through collisions.  When
the fractional ionization is high, neutrals collide frequently with
ions, and the field is coupled tightly to the neutrals.  As the number
of ions drops, they begin to resemble a sieve, and it becomes easier
and easier for neutral gas to slip between adjacent ions, so that the
field decouples from the neutral gas, diffusing through it.  This
process is called ambipolar diffusion, and was first understood to be
astrophysically important by \markcite{ms56} Mestel and Spitzer (1956)
and \markcite{mou79} Mouschovias (1979). In protostellar cores,
ambipolar diffusion of the field out past infalling gas allows mass to
fall in to the central region without dragging field along with it.
In protostellar disks, ambipolar diffusion may allow the magnetic
field to drive jets, reducing the angular momentum of accreting gas.

One of the most directly observable effects of ambipolar diffusion is
to spread out shock waves in dense gas (Draine 1980)\markcite{btd80}.
Normally, the thickness of shocks is determined by gas dynamical or
MHD dissipation in the shock front.  Typical jump or J-shocks in the
interstellar medium have thicknesses much shorter than the cooling
length of the gas, so immediately behind the shock front, the gas gets
quite hot, reaching the post-shock temperature predicted by the shock
jump conditions for adiabatic gas.  On the other hand, if a shock wave
moves faster than the sound speed, $c_s$, but slower than the Alfv\'en
speed in the rarefied ions, $v_{Ai} = |\vec{B}|^2 /
\sqrt{4\pi\rho_i}$, where $\vec{B}$ is the magnetic field and $\rho_i$
is the ion density, a compression wave can travel through the field
and ions, moving faster than the neutral shock front.  The accelerated
ions can then accelerate the neutrals through collisional drag,
spreading the shock front into a broad continuous shock, or C-shock,
in which ambipolar diffusion maintains the thick shock front.
C-shocks typically have thicknesses greater than the cooling length,
so the gas cools as it is accelerated, and never reaches the high
temperatures predicted for J-shocks.  A recent review covering these
points was written by \markcite{dm93} Draine \& McKee (1993).

C-shocks were shown to be unstable by Wardle (1990, 1991a,b)
\markcite{war90}. \markcite{war91a} \markcite{war91b}
The mechanism depends on the field lines in a steady-state C-shock
being perturbed so as to no longer be exactly perpendicular to the
flow, as shown in Figure~\ref{cartoon}.  The ions have low inertia and
collide frequently with neutrals, so a drag force due to the component
of the neutral velocity parallel to the perturbed field lines drives
the ions along the field lines into the valleys and away from the
peaks of the field lines.  

The density of ions in the valleys depends inversely on the wavelength
of the perturbation for constant perturbation size, for two reasons.
First, the ions have longer to travel to reach a valley at longer
wavelengths, and second, the field lines are more normal to the flow
at longer wavelengths, so the component of the neutral drag force
driving ions into the valleys is lower.  As the ions concentrate in
the valleys, and rarefy at the peaks, the drag of the neutrals on the
ions changes proportional to the ion density, transferring neutral
momentum to the field, and producing a growing perturbation that leads
in turn to more ion accumulation.

There are stabilizing forces at both long and short wavelengths that
act against the perturbation, due to magnetic pressure and tension,
respectively.  At long wavelengths, magnetic pressure acts on field
lines compressed into valleys by the increased drag forces there,
slowing or preventing further growth.  At short wavelengths, magnetic
tension acts against the curvature of field lines in valleys and peaks
caused by increasing perturbation size, again slowing or preventing
further growth.  As a result, there exists an intermediate wavelength
of maximum growth, computed by Wardle (1990) in his linear analysis.  
In three dimensions, the fastest growing modes were found by Wardle to
be slabs, with no variation in the third direction.  If there were
variation, the compressed magnetic fields in the valleys could twist
out sideways into the third dimension, releasing magnetic pressure,
but decreasing the drag forces acting along the field lines, and so
slowing the growth of the instability.

Prior computations including ambipolar diffusion have focused on the
collapse of protostellar cores and on the stability of C-shocks.  The
first dynamical computation to include ambipolar diffusion was done by
\markcite{bs82} Black \& Scott (1982) using a two-dimensional,
first-order code that included an iterative approximation to an
implicit method of computing the ambipolar diffusion.  \markcite{pm83}
Paleologou \& Mouschovias (1983) described a one-dimensional code that
included an adaptive mesh and an implicit method for the ambipolar
diffusion.  \markcite{fm92} Fiedler \& Mous\-chovias (1992) described
a two-dimensional code with adaptive mesh and an implicit method.
Mouschovias \& Morton (1991) \markcite{mm91} computed protostellar
collapse problems by integrating over the vertical structure of an
accretion disk and using an implicit method to model the horizontal,
axisymmetric flow.  T\'oth (1994, 1995) \markcite{tot94}
\markcite{tot95} used a two-dimensional, flux corrected transport code
to investigate C-shock stability, using both explicit and implicit
methods.  Mac Low et al.\ (1995) \markcite{mac95} added ambipolar
diffusion to the ZEUS code\footnote{Available for community use by
registration with the Laboratory for Computational Astrophysics at
lca@ncsa.uiuc.edu} (Stone \& Norman 1992a, b) \markcite{sn92a}
\markcite{sn92b} and investigated its
effects on Balbus-Hawley instabilities as a first application.

In this paper we examine the nonlinear evolution of the Wardle
instability with high-resolution two-dimensional and three-dimensional
computations, extending the work of Mac Low et al.\ (1995) by adding
the ion mass conservation equation, and a semi-implicit treatment of
the ion momentum equation, following T\'oth (1995), in order to
compute two- and three-dimensional models of unstable C-shocks.  A
preliminary version of this paper was presented by Mac Low \& Smith
(1997) \markcite{ms97}. \markcite{jms97} Stone (1997) describes
similar computations.

\section{Scales and Parameters}

The time and length scales for our problem are the
time for material to flow through a C-shock (e.g., T\'oth 1995;
Wardle 1990), given by the time required for every neutral to
collide with an ion $\tf = t_{ni} = 1 / \gamma \rho_{i0}$,
where the variables are defined below, under equation~(\ref{mntmi}); and
the thickness of the C-shock, given by $L_{shk} = \sqrt{2}
v_{An} \tf$.

The time-dependent shock flow pattern is fully determined by four
control parameters: the Alfv\'en number \An, the Mach number $M$, the
ion fraction and the initial field orientation.  The Wardle
instability is insensitive to the ion fraction and Mach number so long
as they allow for the existence of a C-shock at all.  Therefore only
one parameter, \An, is necessary to describe the whole class of
transverse, low-ionisation, cold flows: a single flow simulation with
a fixed Alfv\'en number is relevant to a wide range of conditions.

The time scales for ionization and recombination can be important as
well.  Specifically, if the recombination timescale is significantly
shorter than the flow timescale, the possibility exists of the
instability being suppressed.  If the ions are primarily molecular
ions such as H$_3$O$^+$, dissociative recombination rates may be high
enough for this to be relevant (Flower \& Pineau de For\^ets, personal
communication).  We will only consider this case in the extreme limit
of constant ion density, which we will show does indeed suppress the
instability.

In this paper we discuss both two- and three-di\-men\-sion\-al models.
Our ``standard model'' has neutral Alfv\'en number $\An = (v_s/B_0)
\sqrt{4\pi \rho_{n0}} = 25$, initial number density of H atoms $n_{H} =
10^5$ \cc, (note that $\rho_n = 1.4 n_H$), fractional ionization
$\chi = 10^{-4}$, shock velocity $v_s = 5$ \kms, sound speed $c_s =
0.01$ \kms, ion mass $m_i = 10 m_H$, and neutral mass $m_n = 7/3 m_H$,
assuming 10\% He.  We use an ion-neutral collisional coupling constant
$\gamma = <\sigma w_0>/(m_i + m_n) = 9.21 \times 10^{13}$, where we
take $\sigma = 10^{-15}$ cm$^2$, and an effective velocity of $w_0 =
1.9 \times 10^6$ cm s$^{-1}$.  We quote our results in terms of the
length, time, and velocity scales given above, however, as this run
will give results appropriate for $\An = 25$ shocks in a wide variety
of conditions.

The resolution of our runs can best be expressed by giving the number
of zones in the unperturbed shock thickness, $L_{shk}$.  As shorthand,
we use the format L$nn$, where $nn$ are the number of zones.  We have
performed runs with resolutions ranging from L60 to L240 (note that
Stone [1997] used a standard resolution of L21, though he focussed on
slower shocks that had longer wavelengths of maximum growth, as we
discuss in \S~\ref{machsect}).  For our two-di\-men\-sion\-al cases we
used grids ranging from $160 \times 40$ to $640 \times 160$ zones,
while for the three-dimensional cases we used grids with $160 \times 40
\times 40$ and $267 \times 80 \times 40$ zones. 


\section{Numerical Methods}
Mac Low et al.\ (1995) described the basic interface with the ZEUS
code.  Summarizing, that work made four approximations: isothermality
of ions, electrons and neutrals, no electron-ion drift, ion density
dependent in power-law fashion on neutral density, and no ion inertia
or pressure.  This allowed us to neglect, respectively, the energy
equation, Ohmic diffusion, and the equations of ion mass and momentum
conservation.  This approach has proved adequate for modelling
protostellar disks, but broke down for both physical and numerical
reasons when applied to C-shocks.  The Wardle instability depends on
the flow of ions along buckling field lines in the shock front, and so
was suppressed by the neglect of the ion mass conservation equation.

Neglect of ion inertia and pressure had allowed the replacement of the
ion momentum conservation equation by an algebraic equation expressing
the balance between Lorentz forces and ion-neutral drag in determining
the drift velocity between ions and neutrals.  This approach is
physically accurate and allows time steps determined by the equivalent
of the Courant condition for ambipolar diffusion, $ \Delta t \le \pi
\gamma \rho_i \rho_n (\Delta x)^2 / |\vecb|^2 $ (Mac Low et
al. 1995). However, both T\'oth (1994) and we have found that this
approach is numerically unstable in the presence of steep velocity
gradients as in C-shocks.  T\'oth (1994) found that even an explicit
treatment of the ion momentum equation was insufficient to entirely
suppress numerical instability; T\'oth (1995) showed that a
semi-implicit treatment of this equation was far more stable, an
approach we have followed with some modifications described below.

The equations we solve in the current version of the code are then the
neutral and ion continuity equations, the induction equation, and the
neutral and ion momentum equations:
\begin{eqnarray}
\partial \rho_n/\partial t &=& - \nabla \cdot (\rho_n \vecv_n) \\
\partial \rho_i / \partial t &=& - \nabla \cdot (\rho_i \vecv_i) \\
\rho_n  (\partial \vecv_n / \partial t) &=& - \rho_n (\vecv_n \cdot \nabla)
   \vecv_n - \nabla P_n \nonumber \\
 & & +\gamma \rho_i \rho_n (\vecv_i - \vecv_n) 
   \label{mntmn} \\
\rho_i (\partial \vecv_i / \partial t) &=& - \rho_i (\vecv_i \cdot \nabla)
   \vecv_i - \nabla P_i + \gamma \rho_i \rho_n (\vecv_n - \vecv_i) \nonumber \\
 & &   + (1 / 4\pi) (\nabla \times \vecb) \times \vecb
   \label{mntmi}       \\
(\partial \vecb / \partial t)&=&\nabla \times (\vecv_i \times \vecb) 
    \label{induct} \\
\nabla \cdot \vecb&=&0
\end{eqnarray}
where the subscripts $i$ and $n$ refer to the ions and neutrals, $\rho$,
$\vecv$, and $P$ are density, velocity and pressure for each fluid, 
$\vecb$ is the magnetic field, and $\gamma$ is the collisional 
coupling constant between the ions and neutrals. 

We treat the ions as a separate fluid in the code, using the standard
ZEUS algorithms to update them, except in the momentum equation where
we make several changes to implement the semi-implicit algorithm.  The
basic idea of this new algorithm is to treat the stiffest terms in the
equation implicitly, but maintain the standard operator splitting for
the other terms.  T\'oth (1995) included the drag term and the
transported fluxes in his semi-implicit computation.  We find that we
also needed to include the magnetic pressure forces in order to
maintain balance between them and the drag terms.

We rewrite equations~(\ref{mntmn}) and~(\ref{mntmi}) in conservative
form using momenta $\vec{s} = \rho \vecv$, and momentum fluxes $F_n$
and $F_i$, and place them in finite difference form.  We then
get two equations in the two unknowns $\vec{s}_n^{n+1}$ and
$\vec{s}_i^{n+1}$ for our implicit system: 
\begin{eqnarray}
\frac{\vec{s}_n^{n+1}V^{n+1} - \vec{s}_n^n V^n}{\Delta t} & = & -\nabla F_n \\ 
 &+&\gamma V^{n+1}(\rho_n \vec{s}_i^{n+1} - \rho_i \vec{s}_n^{n+1}),
 \nonumber \\
\frac{\vec{s}_i^{n+1}V^{n+1} - \vec{s}_i^n V^n}{\Delta t} & = & - \nabla (F_i +
    \frac{\vecb^2}{8 \pi}) \\
 &+&\gamma V^{n+1}(\rho_i \vec{s}_n^{n+1} - \rho_n \vec{s}_i^{n+1}) \nonumber,
\end{eqnarray}
where time levels are given by superscripts, the finite difference
time step is $\Delta t$, and the inclusion of zone volumes $V$ allows
for a moving grid.  We find the semi-implicit
update to the momenta by solving these equations for $\vec{s}_n^{n+1}$
and $\vec{s}_i^{n+1}$:
\begin{eqnarray}
\vec{s}_n^{n+1} &=& \\
  &  &\left\{ (\vec{s}_n^n V^n - \nabla F_n \Delta t)(1 + \gamma \Delta t
     \rho_n) \frac{\mbox{}}{\mbox{}}\right. \nonumber \\
  &+& \left. \gamma \Delta t \rho_n (\vec{s}_i^n V^n - \nabla \left[F_i +
     \frac{\vecb^2}{8 \pi}\right] \Delta t)\right\} \nonumber \\
  & &\{ [1 + \gamma \Delta t (\rho_n + \rho_i)] V^{n+1}\}^{-1}, \nonumber \\
\vec{s}_i^{n+1} &=& \\
  &  &\left\{ (\vec{s}_i^n V^n - \nabla \left[F_i + \frac{\vecb^2}{8
     \pi}\right] \Delta t)(1 + \gamma \Delta t \rho_i) \right. \nonumber \\
  &+& \left. \gamma \Delta t \rho_i (\vec{s}_n^n V^n - \nabla F_n
     \Delta t)\frac{\mbox{}}{\mbox{}} \right\}
     \nonumber \\
  & &\{ [1 + \gamma \Delta t (\rho_n + \rho_i)] V^{n+1}\}^{-1}. \nonumber 
\end{eqnarray}
Of course, the magnetic pressure updates must be removed from the
source step where they would normally be performed.

Unfortunately, including the ion momentum equation limits our timestep
to $\Delta t \le \Delta x / v_{Ai}$, and we can no longer use the
substepping scheme described by Mac Low et al.\ (1995).  As a result,
computations through a flow time require many $ \times 10^5$ cycles.
The total number of cycles goes as $\chi^{1/2}$, however, so
increasing the ionization fraction helps, so long as it does not get
so high that no C-shock is possible.  T\'oth (1995) showed that a
faster algorithm could be implemented by going to a fully implicit
method, at the cost of greatly increased code complexity.

For initial conditions we used a numerical solution for an isothermal
C-shock \markcite{sm97} (Smith \& Mac Low 1997) interpolated onto our
grid.  We then added a random zone-to-zone fluctuation to the neutral
velocity. In order to maintain the shock stationary on our grid, we
set up the grid with defined values for the field variables at both
ends---that is with ``inflow'' boundaries, but with the right side
having a negative inflow velocity.  We used periodic boundaries along
the sides of the grid.

We can maintain the steady-state C-shock indefinitely in one dimension
(Smith \& Mac Low 1997).  In two and three dimensions, we can maintain
it for many thousands of cycles until the physical instability sets
in, as shown in Figure~\ref{cshk}.  We also showed that we converge on
the steady-state solution starting from a discontinuous initial
condition in Smith \& Mac Low (1997).

\section{Instability}

\subsubsection{Linear growth}

Let us begin by verifying that we are indeed seeing the instability in
the linear regime.  We follow Stone (1997) in using the magnetic
energy in the field component parallel to the shock velocity, $B_x^2$,
as a diagnostic.  The growth rate depends almost entirely on the
neutral Alfv\'en number of the gas and the angle of the shock to the
field, as discussed by Wardle (1990) and T\'oth (1994).  The analytic
growth rate $s$ can be derived from Figure~7 of Wardle (1990).  Note,
however, that that Figure gives the quantity $s \tf$, and that it
appears to assume the value of \tf\ given in Wardle's
equation~(2.12).  The parameters quoted by Wardle would actually give
a coefficient of 5.4 rather than 5.0, but that does not agree with our
numerics, while 5.0 does.  (Note that, for the parameters we use, the
coefficient is 6.5, but that is due to our different choices of $m_i$
and so forth.)    

In Figure~\ref{linear} we show the parallel field energy as a function
of time for our standard case for numerical resolutions of L60, L120,
and L240.  For this case, we derive a value of $s = 42.5
\tf^{-1}$, using Wardle's value of the coefficient.  In
Figure~\ref{linear} we show curves corresponding to $s=(41, 42.5, 44)
\tf^{-1}$, showing the excellent agreement with the analytic
value.  We have also measured the growth rate for our $\An = 10$ run,
and get equally good agreement with the analytic value of $s = 4.2
\tf^{-1}$.

\subsubsection{Nonlinear Development}
\label{nonlinsect}

The striking feature of the nonlinear development is the formation of
long, thin filaments of dense gas.  These filaments begin to merge
with each other so that the wavelength of the instability grows as it
becomes increasingly nonlinear.  (The nonlinear Rayleigh-Taylor
instability also develops merging filaments, as described by Read
[1984] \markcite{rea84} and Youngs [1984] \markcite{you84}).  We show
this development in Figure~\ref{time}.

In our standard model, numerical instabilities prevent computation of
the further development of the filaments after the last time shown in
Figure~\ref{time}.  However, we have been able to extend our
computations of lower neutral Alfv\'en number shocks to later times,
as shown in Figure~\ref{time10}.  We confirm the finding of Stone
(1997) that, in the nonlinear stage when the distance between
filaments reaches the shock thickness, a new filament appears between
the existing ones.

The thickness of the region between the initial ion deceleration zone and
the final post-shock region increases with time as shown in
Figure~\ref{thk}.  The thickness grows with a velocity just
equal to the post-shock velocity of the shock.  It appears that once
the instability is established, it does not relax back to a uniform
medium easily, if at all.  The post-shock state is thus a tree of
merging filaments (or in three dimensions, merging sheets, as we
discuss below.)

We find that the filaments' width drops to the single zone level at
every resolution that we have run the problem.  Stone (1997) suggests
that gas pressure eventually limits their thickness, but points out
that the resolution needed to see this effect would be enormous,
requiring a dynamic range over $10^4$.  We have confirmed this
empirically by running several different Mach 25 runs with sound
speeds as high as $0.4v_{An}$ at our highest resolution, with no
discernible difference in their evolution.  However, as we show next,
the filaments are probably manifestations of the Brandenburg \&
\markcite{bz94} \markcite{bz95} Zweibel (1994, 1995; hereafter BZ)
ambipolar diffusion singularity, which they showed in the simplest
case overwhelms thermal pressure, and so may well become
extraordinarily thin.

\subsection{Ambipolar Diffusion Singularity}

In Figure~\ref{bfld} we show the components of the magnetic field
parallel and perpendicular to the shock velocity, along with the
magnetic pressure.  Although the perpendicular component remains
dominant, as can be seen in the magnetic pressure plot, the parallel
component undergoes a sharp reversal at each filament.  A profile of
the parallel field is shown in Figure~\ref{bprof}. The field lines
thus are basically perpendicular, but with extremely sharp kinks in
them at each filament.  The reversal in the parallel component appears
to be enough to drive the BZ
\markcite{bz94} \markcite{bz95} ambipolar diffusion singularity.  This
singularity steepens magnetic pressure gradients in the presence of
ambipolar diffusion, due to the nonlinear nature of the ambipolar
diffusion.

The actual field behavior that we compute can be understood
analytically.  The simulations demonstrate that the magnetic field
parallel to the flow, B$_x$, grows exponentially (Fig.~\ref{linear}),
evolving towards a saw-tooth structure as the filaments become
thinner. That is, B$_x$ becomes a linear function of y. In contrast,
the ambipolar diffusion analysis of Brandenburg and Zweibel (1994)
delivers a relation B$_x$ $\propto y^{1/3}$. Their analysis, however,
assumed the magnetic field to be zero within the filament whereas here
we have the opposite case: the field B$_y$ dominates. Hence to
understand these different relationships we undertake a more general
analysis.

The aim, following BZ, is to search for
hydrostatic steady flow solutions to the induction equation.  To
begin we take the streaming (or drift) velocity from equation~(\ref{mntmi})
\begin{equation}
\vecv_D =
 (\vecv_n - \vecv_i) =
- (1 / 4\pi\gamma \rho_i \rho_n )(\nabla \times \vecb) \times \vecb  
   \label{vdrift}
\end{equation}
to eliminate the ion velocity from the induction equation~(\ref{induct}):
\begin{equation}
(\partial \vecb / \partial t) = \nabla \times (\vecv_n \times \vecb) +
(1 / 4\pi\gamma \rho_i \rho_n ) \nabla \times \left[\{(\nabla \times
\vecb) \times \vecb\} \times \vecb\right].  \label{induct1}
\end{equation} 
We take a uniform flow in the x-direction i.e. $B_x(y)$, $v_{xn}(y)$
and $B_y = B_0$, a constant as required for vanishing divergence.
Ignoring field amplification due to the first shear term (that is,
taking $v_{xn}$ constant), the induction equation becomes
\begin{equation}
4\pi\gamma \rho_i \rho_n \frac{dB_x}{dt} = \left(B_0^2 +
B_x^2\right)\frac{d^2B_x}{dy^2} + 2B_x\left(\frac{dB_x}{dy}\right)^2
\label{induct2}.
\end{equation} 
The solution to the steady state form is
\begin{equation}
y = k_1 + k_2\left[2\tan~\theta + \frac{\tan^3\theta}{\sin^2\theta}\right]
   \label{induct3}
\end{equation}
where the field angle $\theta = \arctan\,(B_x/B_o)$, and $k_1$ and
$k_2$ are integration constants. The second term dominates and yields
the Brandenburg \& Zweibel (1994) solution in the limit $B_0$
$\rightarrow 0$. Such a solution may be relevant to the study of
parallel-field C-shocks that we plan. The first term, however,
dominates for the case studied here, and indeed predicts a linear
$y-B_x$ behaviour.

This singularity steepens magnetic pressure gradients in the presence
of ambipolar diffusion, due to the nonlinear nature of the ambipolar
diffusion.  Among other things, the resulting steep gradients trigger
numerical instabilities, visible as the single-zone regions of opposed
field along the filaments.  These regions produce anomalous fields and
extremely high ion velocities, reducing the timestep to impractically
small values.  Physically, the exceedingly sharp discontinuities in
field direction produced by this singularity will drive strong current
sheets along the filaments, with uncertain consequences probably
including heating and possibly the production of energetic particles.

\subsection{Three-Dimensional Structure}

To study the three-dimensional structure of the instability we
performed two runs at L120.  One was of our standard case, while the
other had a zone-to-zone perturbation $\delta v_{n1}/v_{n1} = 10^{-6}$
rather than 0.01.  We found that the thin filaments seen in the
two-dimensional models developed in the low perturbation run into
coherent mildly rippled layers, as shown in Figure~\ref{3d}.  In the
standard case, on the other hand, we found that each vertical plane
developed essentially independently, resulting in spikes rather than
layers.  This appears to have been an effect of the initial
perturbation, suggesting that the structure in the third dimension in
nature will be determined by the existing perturbation structure in
the pre-shock gas unless it is quite uniform on scales of $L_{shk}$.

\label{3dsec}

\subsection{Resolution Studies}

In order to understand the effects of numerical diffusivity on our
results, we ran our standard case in two dimensions at resolutions of
L60, L120, and L240, and in three dimensions at resolutions of L60 and
L120, as we discussed at the beginning of this section.  We show the
density distribution at the end of the run for these three different
cases in two dimensions in Figure~\ref{res}.  Examining this figure,
it is clear that the filamentary nature of the nonlinear instability
has been well resolved, but that we have not converged on the exact
value of the wavelength of maximum growth.  As was shown in
Figure~\ref{thk}, the thickness of the shocked layer is converged to
within a few percent.  

The thickness of individual filaments in the transverse direction
appears to be determined entirely by the dissipation scale physically,
due to the BZ ambipolar diffusion singularity, as we discussed above,
in \S~\ref{nonlinsect} Effectively, they become infinitely thin.
Indeed, they reach thicknesses of one or two zones at every
resolution, at which point the extreme gradients cause numerical
instabilities.

\section{Variations on the Theme}

\subsection{Mach Number Dependence}
\label{machsect}

In the linear regime, the growth rate and wavelength of maximum growth
depend on the neutral Alfv\'en number, \An, as shown by Wardle (1990),
and confirmed in some detail by T\'oth (1994, 1995), as well as in
Figure~\ref{linear}.  The slower growth rate and longer wavelength of
maximum growth in the shock with $\An = 10$ shown in
Figure~\ref{time10}, as compared with our standard model, agree with
the linear theory.  We find qualitative differences in the nonlinear
development of the instability at different \An\ as well, as can be
seen by comparing Figures~\ref{time} and~\ref{time10}.  At lower
values of \An, the larger wavelength of maximum growth means that
there is more space between the resulting filaments, and each one is
thicker compared to the initial shock thickness.

\subsection{Ionization Equilibrium}

Until now, we have assumed that ionization and recombination
timescales were long compared to the flow time through the shock
$\tf$.  If both ionization and recombination timescales are very
short compared to the flow time, the ion density will remain constant
with time, or vary as a power law of the neutral density, suppressing
the instability.  The version of ZEUS described by Mac Low et
al. (1995) works in this regime, so we ran our standard case with
constant ion density using that version.  We did indeed find that
after $0.4 \tf$ there was no indication of the Wardle
instability.  At that point, however, the run was ended by numerical
instabilities caused by the large ion-neutral streaming velocities,
which caused strong zone-to-zone fluctuations in the computed parallel
velocities.  In Figure~\ref{neut} we compare the neutral density
distribution for our standard initial conditions to the result for
constant ion density, to show the lack of evolution.

\subsection{Initial Perturbations}

The form and strength of the initial perturbations we use to trigger the
instability can make a difference in the results.  In our first runs
we used a 1\% zone-to-zone perturbation in density; later we followed
T\'oth (1995) in using perturbations in parallel neutral velocity $v_{n1}$
instead, with magnitudes of 1\% and 0.0001\%.  We found that 1\%
perturbations were already large enough to significantly change the
final nonlinear results, especially in three dimensions, as we
discussed in \S~\ref{3dsec}.  Even in two dimensions, the larger
perturbations change the wavelength of maximum growth significantly,
as shown in Figure~\ref{pert}, where the nonlinear development of the
standard case is compared to that of a low perturbation case, which develops
a longer wavelength.

\section{Implications for Molecular Hydrogen Emission}

One measurable consequence of the C-shock instability is that the line
emission directly from the shocked layer will differ from the
steady-state case, and should even be a slowly varying function of
time. Warm C-shocks vibrationally excite the hydrogen molecules. We
may thus expect that the fraction of H$_2$ in higher vibrational
levels would decrease with time as frictional heating is spread out
over a thicker layer. To test this, we employ the ``cool C-shock''
approximation, that requires the temperature $T \ll 2m_H v_s^2/k$.
This is valid provided cooling is very efficient so that the thermal
pressure and pressure gradient can be neglected. Then, the isothermal
assumption is not critical to the flow pattern so long as a low
temperature is chosen. In this case, an effective temperature can be
derived at each location by equating the frictional heating to the
molecular and atomic cooling \markcite{sb90a} \markcite{mds93} (see
Smith \& Brand 1990a, Smith 1993).

The cooling functions we use here have the form $
n_n^{1+f}T^\alpha$ erg s$^{-1}$ cm$^{-3}$.  We study two cases: 
the high-density regime with $\rho_n \sim 10^8\,cm^{-3}$, where H$_2$O cooling
dominates, $f = 1$, and $\alpha = 1.2$; and the low density regime with
$\rho_n \sim 10^{4-6}\,cm^{-2}$, where H$_2$ cooling dominates, $f =
0$, and $\alpha = 3.3$ (Smith 1993).  Frictional
heating is $\propto n_n n_i\,v_{in}^2$. This yields $T^\alpha =
{\eta}v_{in}^2n_i/n_n^f$, where we choose the constant $\eta$ so
that the initial C-shock possesses a maximum temperature of 2,500\,K.
The temperature distribution for the two different cooling cases is shown
in Figure~\ref{temp} for the initial state and a time $0.36\tf$,
after the instability has gone nonlinear.

The column density of H$_2$ molecules in an excited state $j$ with energy
level $kT_j$ is given by
\begin{equation} \label{colj}
N_j = \frac14 g_j N({\rm H}_2) \exp(-T_j / T) / Z(T).
\end{equation}
The statistical weights $g_j = 4 \psi (2J_j + 1)$, where $\psi$ is the
fraction of H$_2$ molecules in the ortho state (assumed to be 0.75
here), J$_j$ is the rotational quantum number, and Z(T) is the
partition function
\begin{equation} \label{pfcn}
Z(T) = 6.135 10^{-3}\,T\,\left[1 - \exp(-5850/T)\right]^{-1},
\end{equation}
(Smith, Davis \& Lioure 1997), assuming local thermodynamic
equilibrium.  The intensity of a line is then given by 
\begin{equation} \label{lintens}
L_j = (hc/\lambda_j) N_j A_j,
\end{equation}
where the radiative deexcitation rates $A_j$ (Turner, Kirby-Docken, \&
Dalgarno 1977)\markcite{tkd77}, along with statistical weights,
wavelengths (Black \& Dishoeck 1987)\markcite{bd87}, and excitation
temperatures for a number of important lines are given in Table~1.

\markcite{ns97} Neufeld \& Stone (1997) concluded that the Wardle
instability would not markedly change the emission expected from
steady-state C-shocks.  However, they only considered shocks with
neutral Alfv\'en number $\An = 10$ and $n_0 = 10^5$ \cc, which will be
dominated by molecular hydrogen cooling.  While we agree that the
emission from these shocks is not strongly modified by the Wardle
instability, we now show that shocks in other regions of parameter
space behave much differently.

One common method of deducing molecular excitation is examining the
ratio of the strong H$_2$ lines 1-0 S(1) and 2-1 S(1), which should
increase as a gas cools and the second vibrational level becomes less
populated. We calculate the column density of molecules in each grid
zone, and then sum over zones to find the total number of molecules on
the grid in each upper energy level, N$_j$.  Radiative transfer could
be added in this treatment. However these quadrupole transitions are
expected to be optically thin.  Figure~\ref{ratio} shows the ratio of
line intensities L(1-0\,S(1))/L(2-1\,S(1)), ignoring foreground
extinction, as a function of time and numerical resolution for both
our standard case and our model with $\An = 10$.  We find that the
instability can shift the ratio strongly when H$_2$O cooling dominates
at $\An = 25$, and more weakly in other cases.  We have not
converged on the actual value of the shift at our current numerical
resolution, primarily because the hot filaments seen in
Figure~\ref{temp} are not yet well resolved.  As unincluded physics is
probably also important there, higher resolution runs are not
particularly justified.

A more general method of determining the molecular excitation is to
correlate column densities (which can be derived from line intensities
using equation~\ref{lintens} and Table~1) in a number of lines against
the excitation temperatures of the lines.  Because of the strong
temperature dependence of the column densities in equation~\ref{colj},
we scale the column densities by $g_j\exp( -T_j /2000 K)$, the
temperature dependence of a slab of gas with temperature 2000 K, where
the statistical weights $g_j$ and excitation temperatures $T_j$ for
important lines are given in Table~1.  In Figure~\ref{excite}, we show
the resulting time-dependent development of the excitation conditions
for our standard case with either H$_2$O and H$_2$ cooling dominant.  
The instability forms both hot and cold regions compared to the
steady-state C-shock, and so produces a marked shift in the integrated
excitation conditions across the shock.

Finally we can examine the detailed line profiles produced by stable
and unstable C-shocks.  We compute them by binning the grid into
angle-dependent velocity bins, and then adding up the intensity of
emission from all zones in each bin (scaled by the square of the
zone-size for runs of different linear resolution) to get a spectrum.
We smooth the resulting profiles with a 1 km s$^{-1}$ Gaussian (except
for the narrow profile from a line of sight perpendicular to the shock
velocity, which we smooth with an 0.1 km s$^{-1}$ Gaussian). The
intensity scales displayed are arbitrary, but consistent across all
plots to allow comparison between lines and angles.

We first examine the effects of angle and numerical resolution on
profiles of the bright 1-0 S(1) line in Figure~\ref{angle}.  We show
the line profile for lines of sight parallel to the shock velocity, at
$45^{\circ}$ to it, and perpendicular to it, for shocks dominated by
H$_2$O and H$_2$ cooling.  Note that the velocity scale is different
for each angle in the figure.  We find that the line is broadened to
the Alfv\'en velocity even on lines of sight normal to the shock
velocity.  We find that the details of our profiles are dependent on
our numerical resolution, as can be seen by comparing different line
profiles in each plot, which are from runs differing by a factor of
two in linear resolution.  However, the gross shape and intensity of
each line appears well described.  The smaller scale variation is due
to the dependence of the results on the details of the filament
structure where most of the hot gas is produced.  Again, the structure
of these filaments is also liable to be influenced by physics not yet
included in our model.

Finally, we examine the profiles of a number of lines observable with
ISO (Figure~\ref{iso}), in the K-band, or with NICMOS on the {\em
Hubble Space Telescope} (Figure~\ref{hstk}), for cooling dominated
either by H$_2$ or by H$_2$O.  We show in grey the profiles predicted
for a steady-state C-shock viewed parallel to the shock velocity.
These are very well resolved, and represent the first publication of
predicted line profiles for most of these lines; they agree well with
the general predictions of Smith \& Brand (1990b) \markcite{sb90b}.  In
black we show the same profiles at the end of our standard run,
showing the shifts in line profile and intensity due to the growth of
hot and cold regions due to the instability.  We also give the derived
total column densities from these lines in Table~1 to allow
computation of arbitrary line ratios.  The convergence of the
unstable profiles can be judged by comparison with Figure~\ref{angle}.
The general direction and size of the shift from the steady-state
value is probably correct, but the quantitative results are probably
good to no better than 20\%.  The general trend we observe is a
strengthening of lines at the soft and hard ends of the spectrum due
to the broader range of temperatures in the C-shock, as well as
changes in the velocity structure due to the filaments.

\section{Summary}

\begin{itemize}

\item The fastest growing modes of the Wardle (1990) C-shock
instability are layers with thickness of order $0.1 L_{shk}$ in
the plane parallel to the field and the shock normal, in agreement
with Wardle's linear analysis.  As predicted, our standard shock with
neutral Alfv\'en number $\An = 25$ goes unstable quite rapidly,
becoming very non-linear in a fraction of $\tf$.

\item When the shock goes non-linear, the layers collapse into thin,
dense filaments (sheets in three dimensions) with density more than a
factor of two above the expected post-shock density.  A dense sheet
also forms across the shock front at the downstream end of the
filaments.  These filaments are physically probabably too thin to be
well-resolved numerically.  T\'oth (1994, 1995) did not see these
filaments, possibly due to the diffusivity of his code, but Stone
(1997) finds a similar morphology.

\item As the instability becomes more nonlinear, the sheets collapse
toward each other, as seen in the upper panels of Figure~\ref{time}.
Thus, longer wavelength modes become dominant as time goes on.  This
process has proceeded in some of our runs until only one wavelength is
left on the grid.  Rayleigh-Taylor instabilities undergo a similar
lengthening of the dominant mode in the non-linear regime.  New
filaments appear when the distance between filaments becomes
comparable to shock thickness.  

\item The thickness of the shock front
increases linearly as the filaments grow, with the ion velocity
dropping sharply at the ends of the growing filaments.

\item The \markcite{bz94}\markcite{bz95} Brandenburg \& Zweibel (1994,
1995) ambipolar diffusion singularity steepens the fields along the
filaments, forcing the filaments to be ever thinner, until the code
breaks down due to the resulting sharp gradients.  Strong current
sheets form along the filaments as a result, with consequences that
probably include heating and ionization, and possibly the production
of energetic particles.  The physics of these filaments have not been
completely captured with our approximations. 

\item We compare the integrated excitation conditions across stable
and unstable C-shocks, giving comparisons of scaled column density to
line excitation temperature, line profiles, and line ratios for our
standard case.  These results are sensitive to the as yet unmodeled
physics of the filaments, as well as numerical resolution of the very
thin filaments, but probably give a generally correct description of
the effect of the instability on observables from C-shocks.

\item Significant variations in emission line properties are predicted
on timescales of fractions of $\tf = (30/n_i)$ yr, where $n_i$ is
the ion number density.  Thus, multiple observations of suspected
C-shocks in dense regions should prove rewarding.

\end{itemize}

\acknowledgments We thank J. Stone for generously sharing his results
prior to submission, S. Beckwith for emphasizing the need for
predictions of multiple line properties, and E. Zweibel for
discussions of the ambipolar diffusion singularity.  MDS thanks the
DFG for financial support.  Computations were performed at the MPG
Rechenzentrum Garching.

\clearpage
\begin{deluxetable}{lllllllll}
\tablewidth{417.24664pt}    
\tablecaption{Representative Molecular Hydrogen Lines}
      \label{linetable}
\tablecolumns{9}

 \tablehead{
 & & & & &  \multicolumn{2}{c}{H$_2$O cooling\tablenotemark{e}} & 
            \multicolumn{2}{c}{H$_2$ cooling                  } \\ 
\cline{6-7} \cline{8-9}
Line & $\lambda$\tablenotemark{a} &  $T_j$\tablenotemark{b} & 
       $A_j$\tablenotemark{c}\mbox{  }     &  $g_j$\tablenotemark{d} &  
       $ t=0$ & $0.36 \tf$ \mbox{  }  & $t=0$ & $0.36 \tf$ 
}

\startdata

2-0 S(5) &  1.0848 &  15763. &   3.28  &  45.      &  3.1e1 &  2.4e1 &  1.5e1 &  2.9e1 \nl   
1-0 S(3) &  1.9570 &  8365.  &  4.21  &  33.       &  6.3e2 &  3.9e2 &  3.1e2 &  1.9e2 \nl   
2-1 S(4) & 2.0035  & 14764. &   5.57  &  13.       &  1.4e1 &  9.9e0 &  6.9e0 &  1.1e1 \nl   
1-0 S(2) &  2.0332  &  7584. &   3.98  &   9.      &  2.5e2 &  1.6e2 &  1.2e2 &  7.1e1 \nl   
2-1 S(3) &  2.0729  & 13890.  &  5.77 &   33.      &  5.2e1 &  3.5e1 &  2.6e1 &  3.3e1 \nl   
1-0 S(1) &  2.1213 &  6951. &   3.47 &   21.       &  7.9e2 &  5.0e2 &  3.9e2 &  2.2e2 \nl   
2-1 S(2) &  2.1536 &  13150.  &  5.60  &   9.      &  2.0e1 &  1.3e1 &  9.7e0 &  1.1e1 \nl   
3-2 S(3) &  2.2008 &  19086. &   5.63  &  33.      &  5.5e0 &  5.3e0 &  2.7e0 &  1.1e1 \nl   
1-0 S(0) &  2.2226  &  6471. &   2.53  &   5.      &  2.4e2 &  1.5e2 &  1.2e2 &  6.5e1 \nl   
2-1 S(1) &  2.2471 &  12550. &   4.98 &   21.      &  6.0e1 &  3.8e1 &  3.0e1 &  3.0e1 \nl   
4-3 S(3) &  2.3439  & 23956.  &  4.58  &  33.      &  6.9e-1 &  1.1e0 &  3.3e-1 &  5.1e0 \nl 
3-2 S(1) &  2.3858 &  17818.  &  5.14  &  21.      &  6.0e0 &  5.2e0 &  2.9e0 &  8.9e0 \tablebreak 
0-0 S(9) &  4.6950 &  10265.   &  4.90   &    69.  &  5.5e2 &  3.4e2 &  2.7e2 &  2.0e2 \nl   
0-0 S(8) &  5.0530 &   8682.  &   3.24    &   21.  &  3.5e2 &  2.1e2 &  1.7e2 &  1.1e2 \nl   
0-0 S(7) &  5.5110  &  7199.   &  2.00   &    57.  &  1.9e3 &  1.2e3 &  9.4e2 &  5.3e2 \nl   
0-0 S(6) &  6.1090 &   5833.  &   1.14  &     17.  &  1.1e3 &  7.4e2 &  5.4e2 &  3.0e2 \nl   
0-0 S(5) &  6.9100  &  4587.  &   0.588   &   45.  &  5.5e3 &  4.1e3 &  2.7e3 &  1.5e3 \nl   
0-0 S(4) &  8.0250 &   3476.  &   0.264  &    13.  &  2.9e3 &  2.4e3 &  1.4e3 &  8.3e2 \nl   
0-0 S(3) &  9.6650  &  2504.   &  0.0984   &  33.  &  1.4e4 &  1.3e4 &  6.5e3 &  4.1e3 \nl   
0-0 S(2) & 12.2790  &  1682.  &   0.0276   &   9.  &  6.6e3 &  7.6e3 &  3.1e3 &  2.3e3 \nl   
0-0 S(1) & 17.0350  &  1015. &    0.00476 &   21.  &  2.9e4 &  4.1e4 &  1.3e4 &  1.2e4 \nl   

\enddata

\tablenotetext{a}{Wavelength in $\mu$m}
\tablenotetext{b}{Excitation temperature in K}
\tablenotetext{c}{Einstein A-values for radiative deexcitation in
$10^{-7}$ s$^{-1}$}
\tablenotetext{d}{Statistical weights}
\tablenotetext{e}{Relative column densities for given times and
cooling, where 1e4$ = 1 \times 10^4$}

\end{deluxetable}

\clearpage
\begin{figure}
\caption[junk]{
The Wardle instability occurs when field lines are perturbed slightly.
The neutral drag forces ions into the valleys, increasing the drag in
the valleys and reducing it at the peaks.  As a result, the valleys
are further deepened, while magnetic pressure expands the peaks as
well, driving an exponential instability.
}
\label{cartoon}
\end{figure}

\begin{figure}
\caption[junk]{ 
Ion (earlier drop) and neutral (later drop) velocities
scaled to the shock velocity.  Dotted lines show the analytic
solution, while solid lines show the two-dimensional higher resolution
(grey) and three-dimensional (black) solutions after $0.11 t_{\rm
flow}$.  The three-dimensional solution has been shifted by $1.5
L_{shk}$ for visibility.}
\label{cshk}
\end{figure}

\begin{figure}
\caption[junk]{
Growth of instability, as indicated by the growth of magnetic energy
in the field component parallel to the shock velocity, $B_x^2$,
normalized to the initial magnetic energy on the grid $B_{y0}^2$.  Our
standard case in two dimensions with neutral Alfv\'en number $\An =
25$ is shown at resolutions of L60 (dotted line), L120 (dashed line),
and L240 (thick solid line).  Overlaid are three exponentials (thin
solid lines) of form $e^{st}$, with $s = (41, 42.5, 44)
\tf^{-1}$, where the analytic value is $s = 42.5 \tf^{-1}$,
as discussed in the text.  In three dimensions we show a run with the
same parameters at a resolution of $L120$ and with smaller initial
perturbations (thick solid line).  The overlaid exponential (thin
solid line) in this case is the analytic value of the growth rate,
normalized appropriately.}
\label{linear}
\end{figure}

\begin{figure}
\caption[junk]{ Time history of (a) neutral parallel velocity, (b)
neutral density, and (c) ion density for
our standard model with a neutral Alfv\'en number $\An = 25$, and
other parameters given in the text, run at a resolution of L240 on a
grid of $640 \times 160$ zones.  }
\label{time}
\end{figure}

\begin{figure}
\caption[junk]{Time history of ion density for our model with neutral
Alfv\'en number $\An = 10$, run at a resolution of L120 on a large
grid of $400 \times 240$ zones.  The first indications of a density
enhancement where another filament will form can be seen in the
last panel, between the upper two filaments.
}
\label{time10}
\end{figure}

\begin{figure}
\caption[junk]{
Shock thickness as a function of time at numerical resolutions of L60
(dotted), L120 (dashed), and L240 (solid), showing the near-linear
growth in the size of the disturbed shocked region.
}
\label{thk}
\end{figure}

\begin{figure}
\caption[junk]{
Parallel and perpendicular components of the field at the end of our
standard run.  Note that the field remains dominated by the
perpendicular component.}
\label{bfld}
\end{figure}

\begin{figure}
\caption[junk]{Profile across filaments of the parallel field
component $B_x$ 
at the end of our standard run, at the midpoint of the grid in the
x-direction.   Values in individual grid zones are shown by triangles.
Note the linear increase in the field, and the extremely sharp
transitions in the filaments.
}
\label{bprof}
\end{figure}

\begin{figure}
\caption[junk]{ 
(a) Comparison of neutral $x$-velocity distribution in the
$x-y$ plane for two- and three-dimensional versions of the standard
model at a resolution of $L_{shk} = 120$ zones (total grid $267 \times
80 \times 40$ zones).  (b) Cuts through the $y-z$ plane of the
three-dimensional model at locations indicated by black lines in (a),
scaled to the same palette.
}
\label{3d}
\end{figure}

\begin{figure}
\caption[junk]{ Comparison of results for runs with resolutions of
L60, L120, and L240.  The log of the neutral density distribution is
shown for three runs with total grids of $160 \times 40$ zones, $320
\times 80$ zones, and $640 \times 160$ zones, for our standard model.
Note the convergence of the morphology and the wavelength of maximum
growth, as well as the consistent increasing trend in the peak
density, which has not yet converged.  }
\label{res}
\end{figure}

\begin{figure}
\caption[junk]{ Comparison of runs with conserved ions versus constant
ion density, showing that if ionization and recombination occur
fast enough to maintain a constant ion density, the instability is
suppressed.  Log of neutral density is shown,
with the same scaling across both images.  The run with conserved ions
is our standard model at a resolution of $L_{shk} = 120$ zones, while
the run with constant ion density was run with ion fraction $\chi =
10^{-6}$ for speed.  (This should make no physical difference, aside
from changing the absolute time and length scales.)  The run at
constant ion density had to be halted due to numerical instability at
this time, but no sign of physical instability was seen.
}
\label{neut}
\end{figure}

\begin{figure}
\caption[junk]{
Density distribution for runs with the
given initial perturbation strength (in both cases, perturbations
consisted of zone-to-zone random changes in the parallel neutral
velocity).  The strong perturbation case is shown at a time $t=0.36
\tf$, while the weak perturbation case is shown at a time $t=0.69
\tf$.  The strong perturbation standard case was only run with
half the number of zones in the y-direction, so it has been doubled
for comparison to the $640 \times 320$ zone weak perturbation case.
}
\label{pert}
\end{figure}

\begin{figure}
\caption[junk]{ 
When drag heating balances radiative
cooling, the equilibrium temperature can be computed assuming that
either (a) water or (b) molecular hydrogen dominates the cooling.  We
show the log of temperature for our standard model before the
instability sets in, and after it has become nonlinear at a time
$0.36\tf$.  The earlier time shows the standard picture of
C-shocks, while the later time emphasizes the difference the
instability makes.  Note the existence of warm gas at the temperatures
predicted by the standard picture, but mixed with colder and hotter
gas.  }
\label{temp}
\end{figure}

\begin{figure}
\caption[junk]{ 
Ratio of emission intensity from the 1--0 S(1) and
2--1 S(1) lines of molecular hydrogen for cooling dominated by (a)
water and (b) molecular hydrogen for our $\An= 25$ run.  Lighter
grey lines show a numerical resolution of L60, darker grey lines are
L120, and black lines are L240.  (c) The same ratio is shown for our
$\An = 10$ run for both molecular hydrogen cooling (dashed) and
water cooling (solid).
}
\label{ratio}
\end{figure}

\begin{figure}
\caption[junk]{ Variation of excitation conditions in an unstable
C-shock at times of 0 (lightest grey), $0.19 \tf$, $0.29
\tf$, and $0.36 \tf$ (black) for cooling dominated by (a)
water and (b) molecular hydrogen.  The column density is scaled
by $g_j\exp (-T_j / 2000\mbox{ K})$, the
functional form of column densities from a uniform slab of gas at a
temperature of 2000 K, where the statistical weights $g_j$ and excitation
temperatures $T_j$ are given in Table~1 for many common lines.  Thus,
such a uniform slab of gas would produce a flat line in this plot.  }
\label{excite}
\end{figure}

\begin{figure}
\caption[junk]{ Variation with angle to the shock velocity of the line
profile from the 1-0~S(1) line of H$_2$ at 2.121 $\mu$m in the final
state of our standard case at resolutions of L60 (lighter grey), L120
(darker grey), and L240 (black), demonstrating the degree of numerical
convergence expected for cooling dominated by (a)
water and (b) molecular hydrogen.  Note that the velocity axes are scaled by
different values at each angle, with $v_s = 25 v_{An}$ for this shock
with $\An = 25$.  }
\label{angle}
\end{figure}

\begin{figure}
\caption[junk]{ A comparison of line profiles from an unstable
C-shock (black) to profiles from a steady-state C-shock with our
standard initial parameters (grey) for a sample of H$_2$ lines
observable by ISO, assuming (a) water cooling and (b) molecular
hydrogen cooling.  The unstable C-shock profiles are computed from the
final state of our highest resolution L240 run.  }
\label{iso}
\end{figure}

\begin{figure}
\caption[junk]{A comparison of line profiles from an unstable C-shock
(black) to profiles from a steady-state C-shock with our standard
initial parameters (grey) for a sample of near-IR H$_2$ lines
observable in the K-band, or with NICMOS on the HST, assuming (a)
water cooling and (b) molecular hydrogen cooling.  The unstable
C-shock profiles are computed from the final state of our highest
resolution L240 run.  }
\label{hstk}
\end{figure}

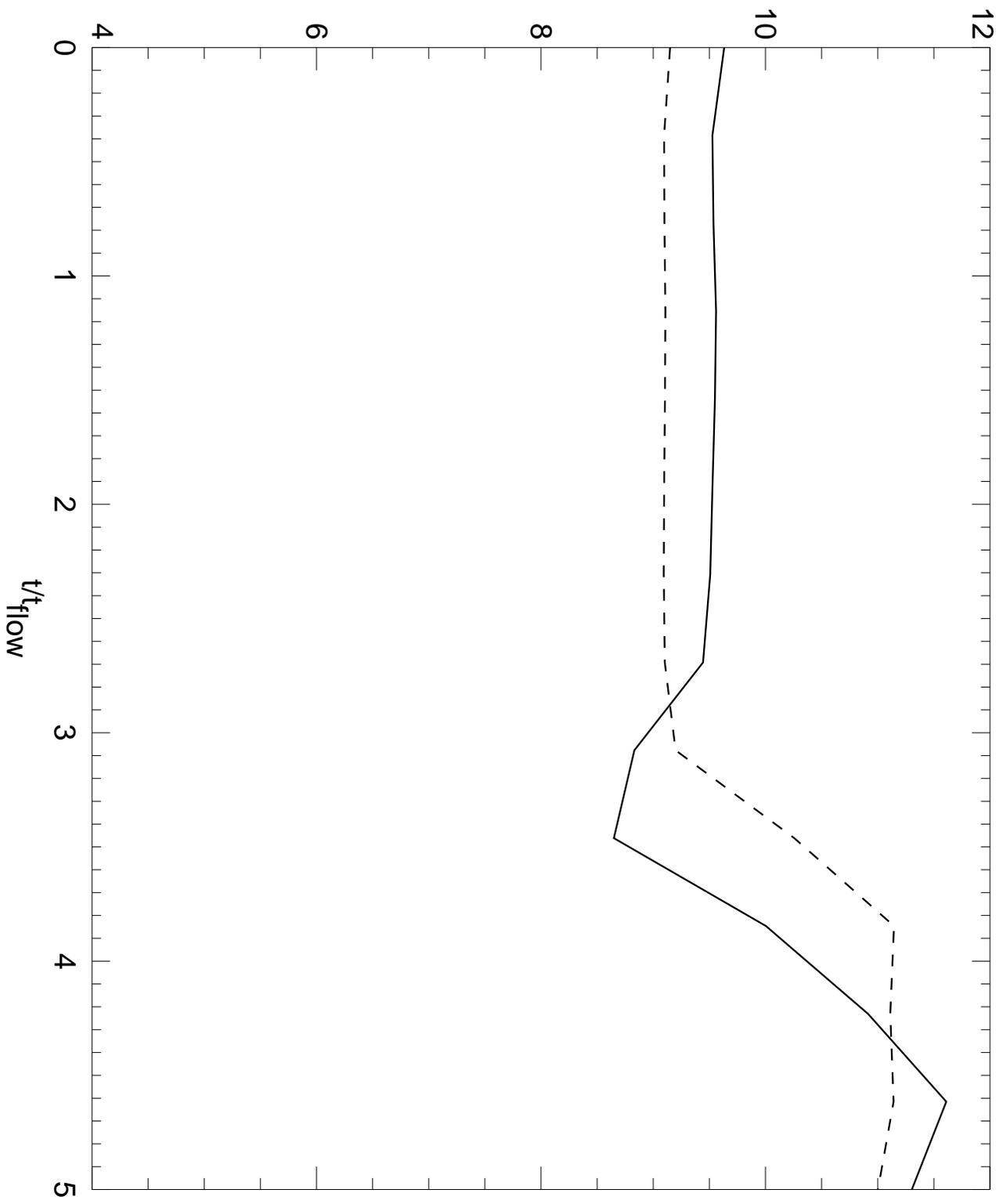
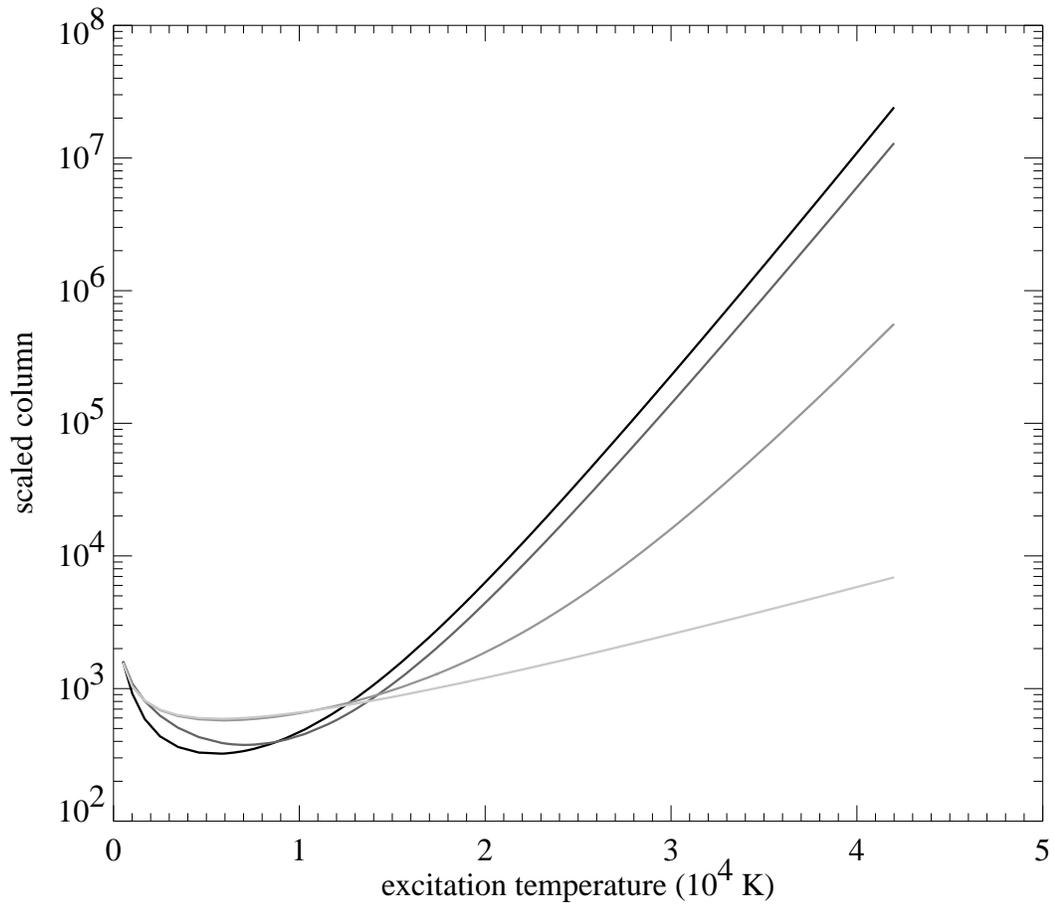
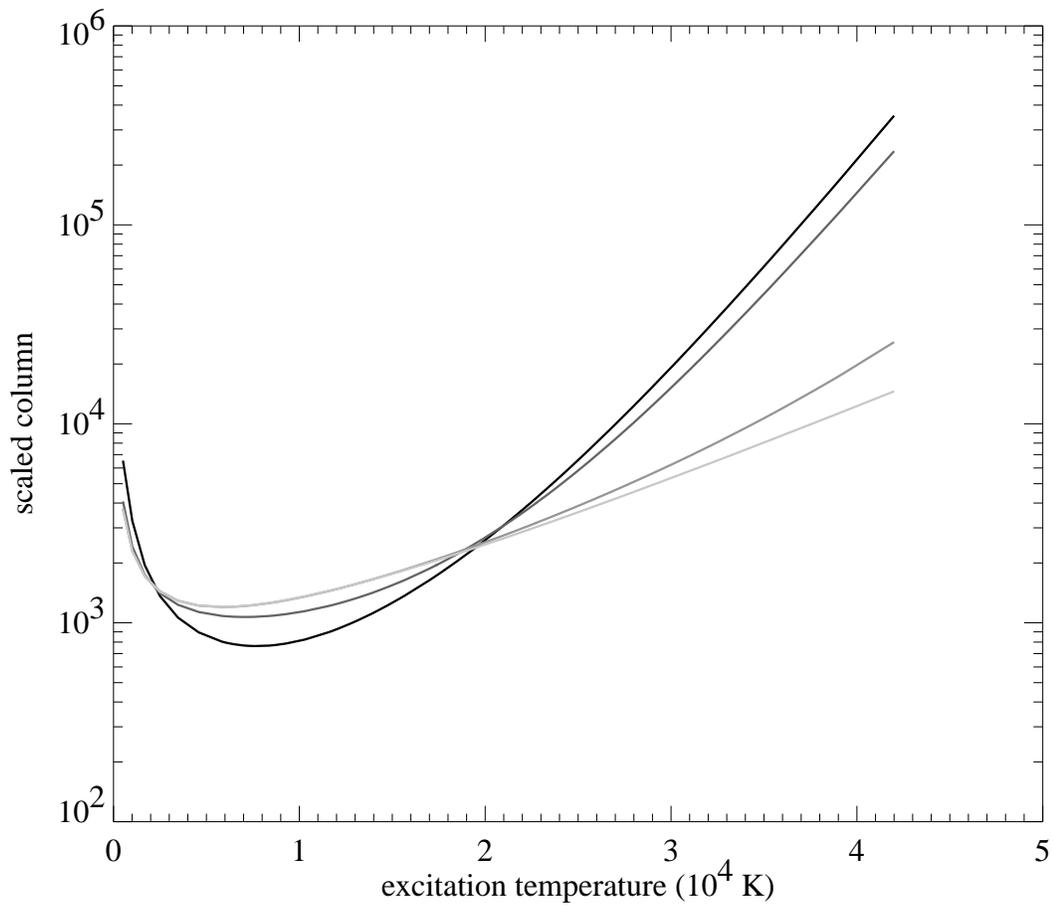
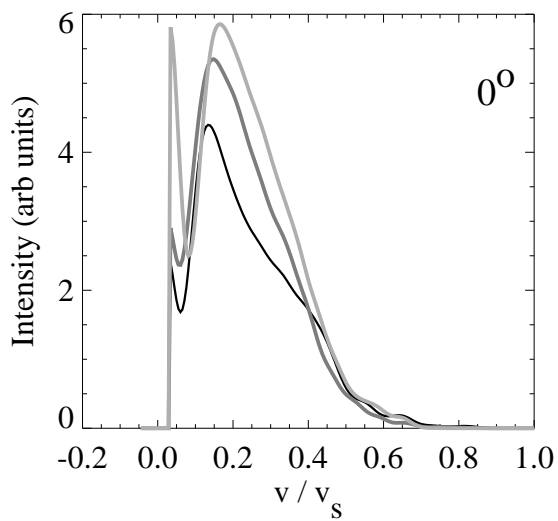
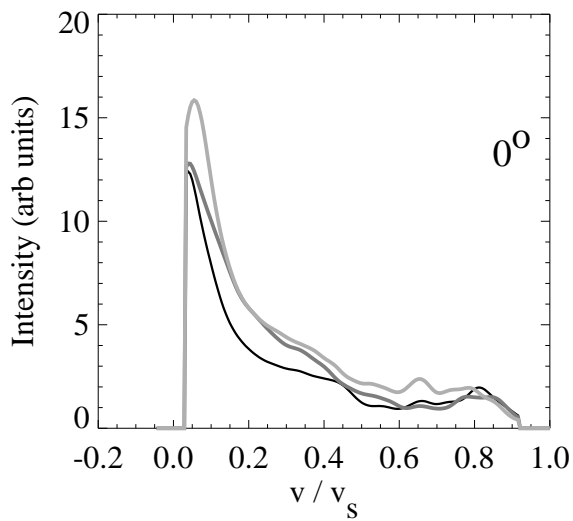
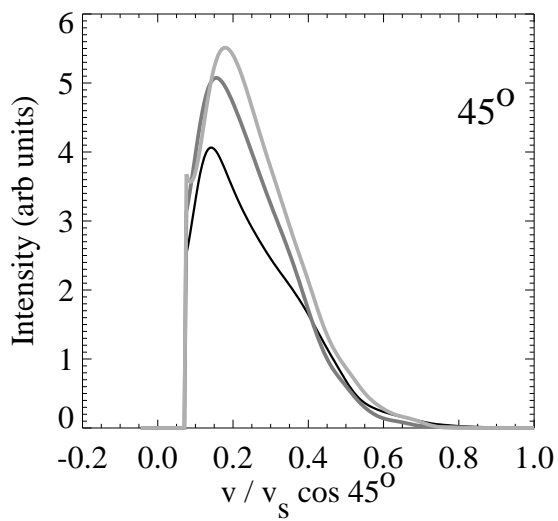
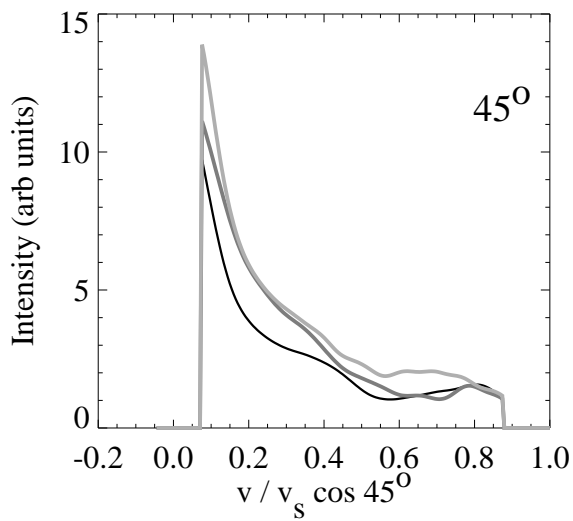
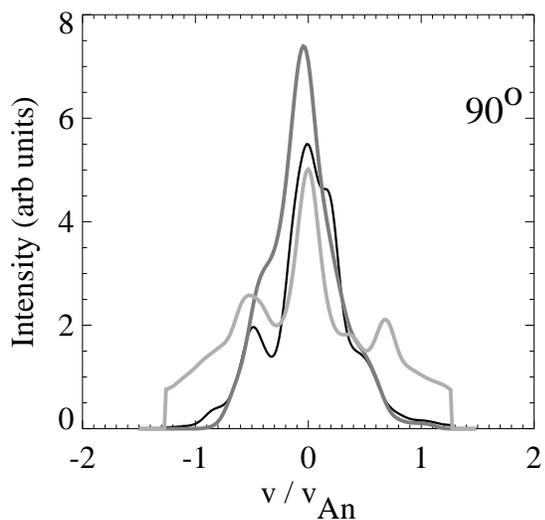
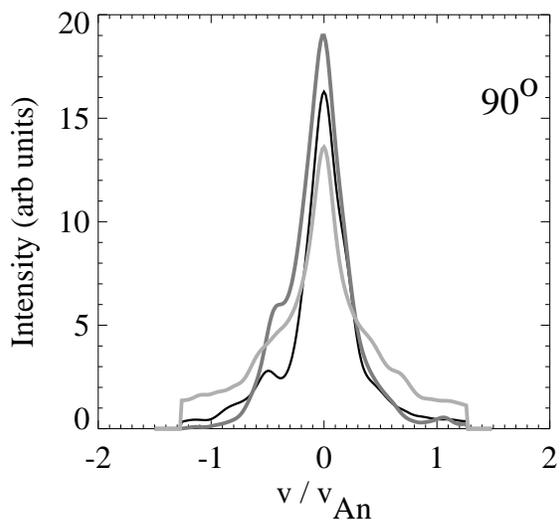
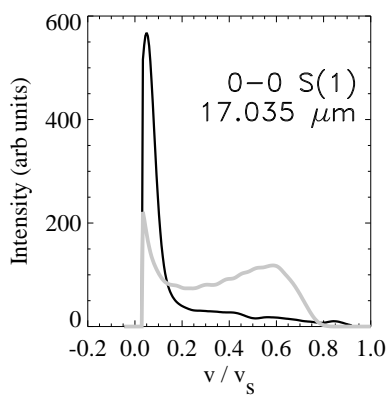
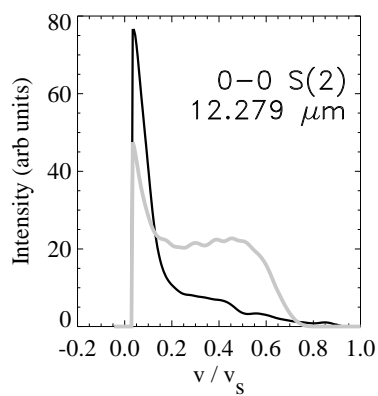
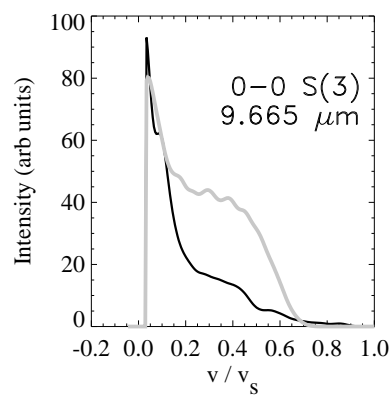
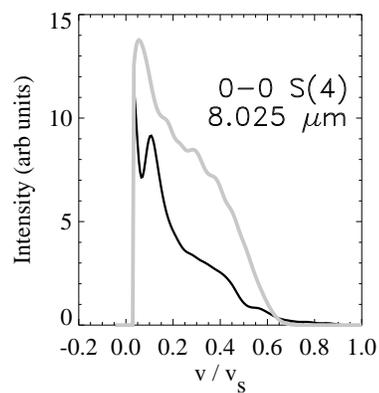
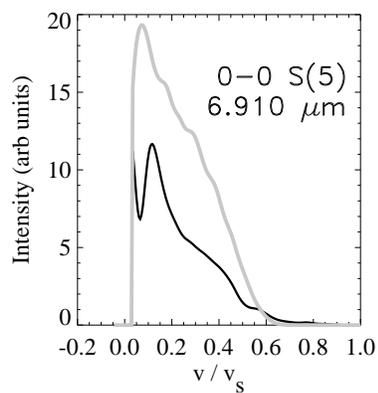
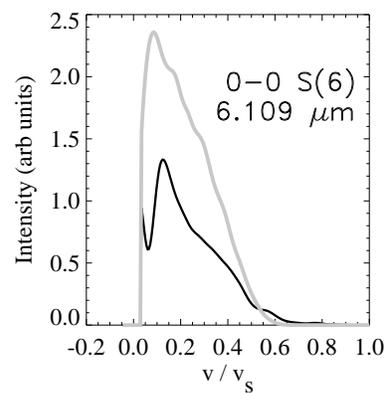
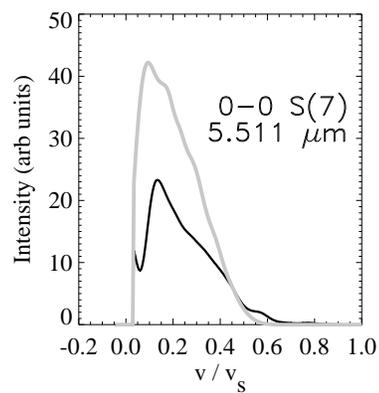
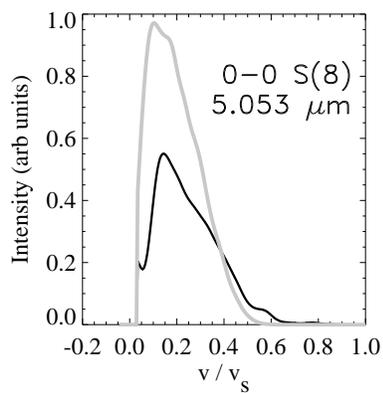
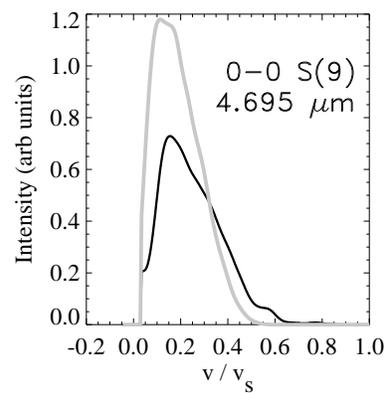
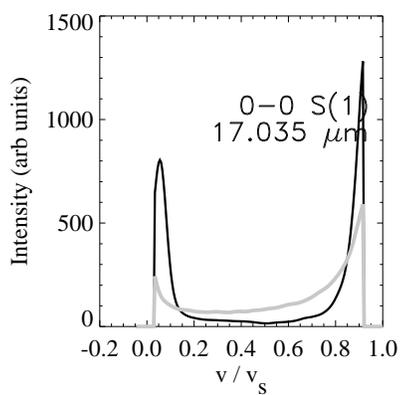
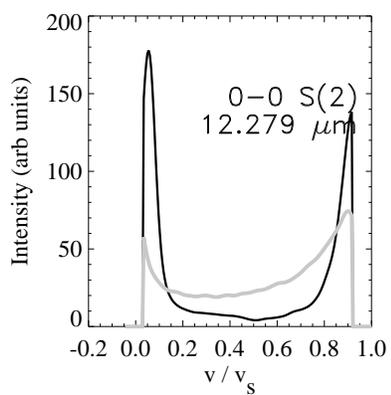
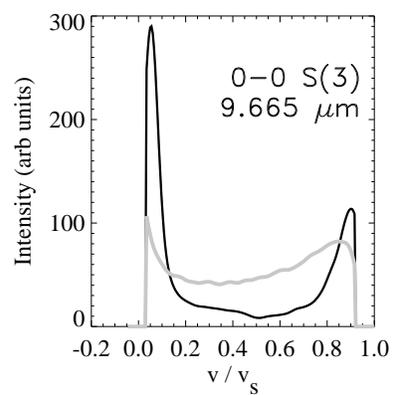
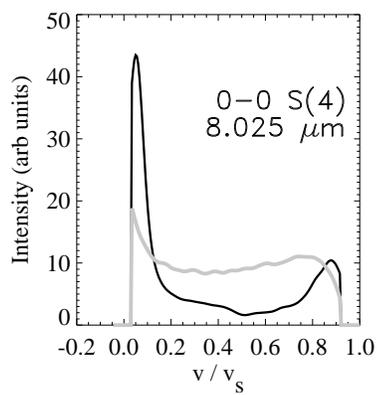
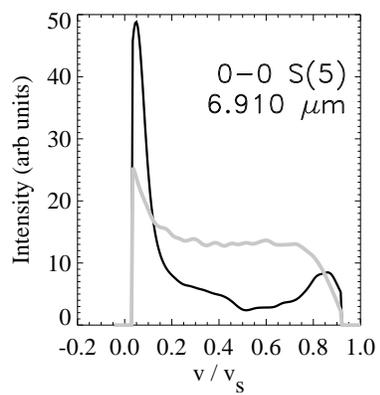
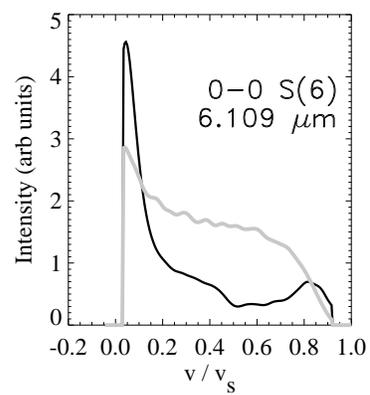
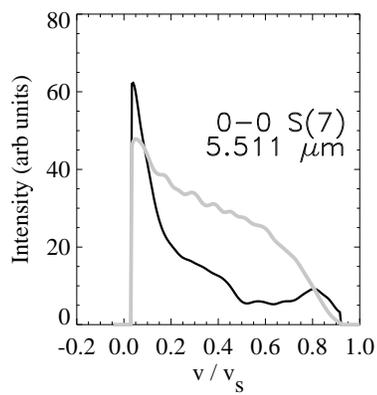
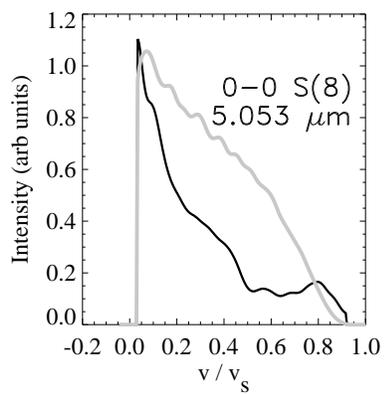
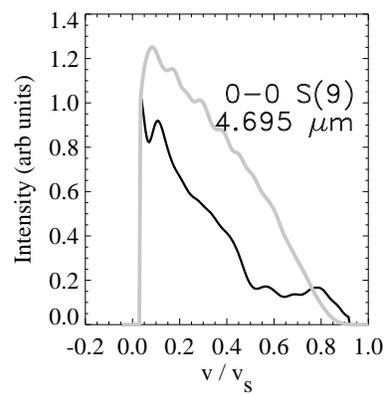
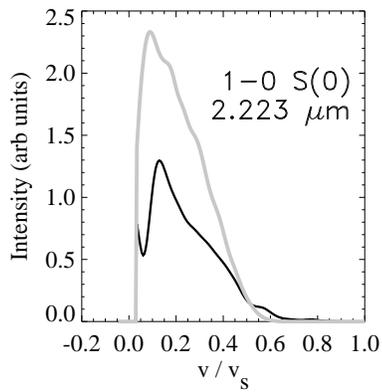
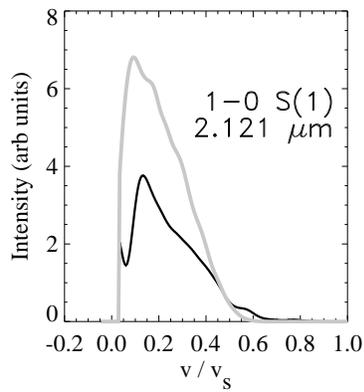
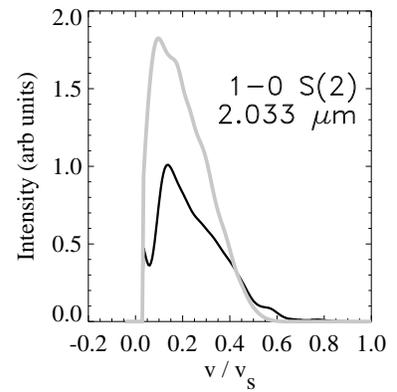
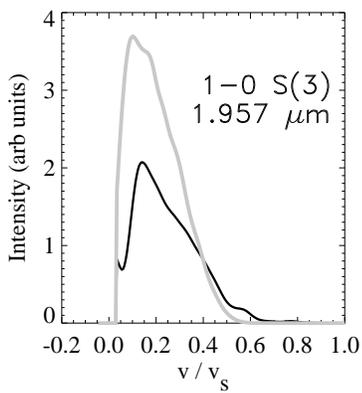
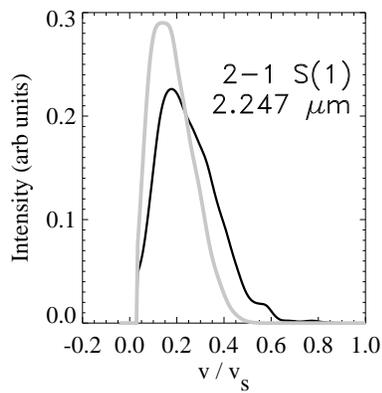
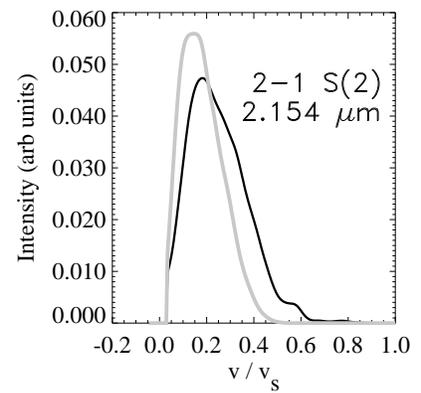
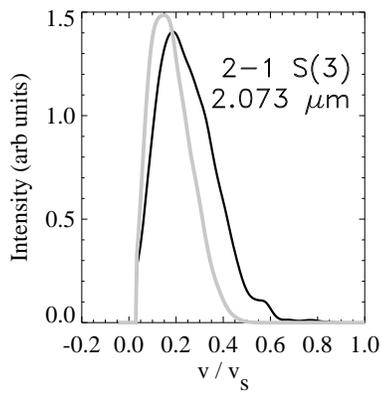
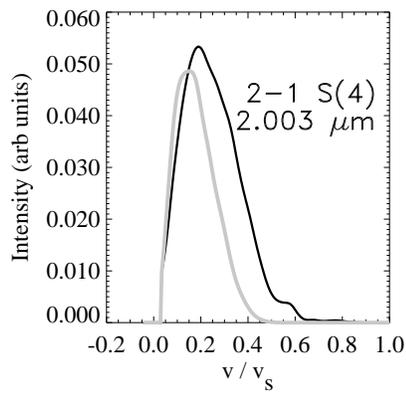
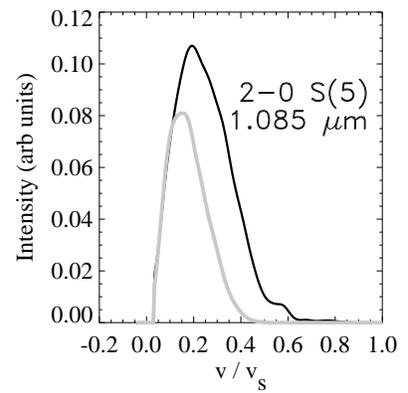
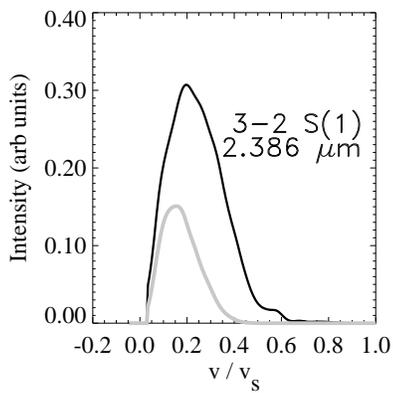
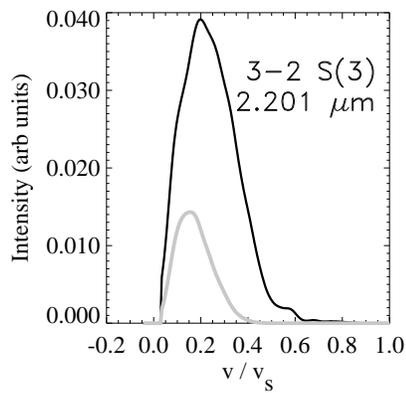
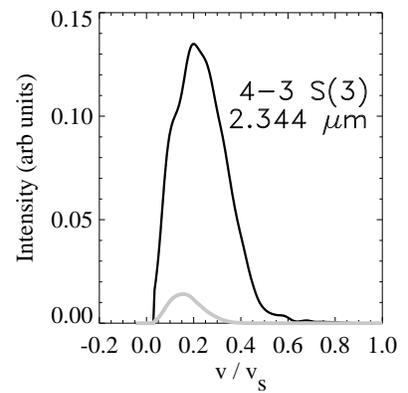
\end{document}